\documentclass[pre,aps,preprintnumbers,amsmath,amssymb,superscriptaddress,twocolumn]{revtex4}
\usepackage{graphicx,subfigure}
\usepackage{pstricks}
\usepackage{xcolor}
\usepackage[breaklinks=true,hidelinks]{hyperref}
\begin{document}

 \title{The emergence of interstellar molecular complexity explained by interacting networks,  
 \\PNAS 119 (30) e2119734119 (2022) \\ \url{https://www.pnas.org/doi/10.1073/pnas.2119734119}}

\author{Miguel Garc\'{\i}a-S\'anchez}
\affiliation{Centro de Astrobiolog\'{\i}a (CSIC-INTA), ctra. de Ajalvir km 4, 28850 Torrej\'on de Ardoz, Madrid, Spain }
\affiliation{Instituto de Investigaci\'on Tecnol\'ogica (IIT), Universidad Pontificia Comillas, 28015 Madrid, Spain}
\affiliation{Grupo Interdisciplinar de Sistemas Complejos (GISC), Madrid, Spain}
 
\author{Izaskun Jim\'enez-Serra} 
\affiliation{Centro de Astrobiolog\'{\i}a (CSIC-INTA), ctra. de Ajalvir km 4, 28850 Torrej\'on de Ardoz, Madrid, Spain }

\author{Fernando Puente-S\'anchez}
\affiliation{Department of Aquatic Sciences and Assessment, Swedish University of Agricultural Sciences, 76551 Uppsala, Sweden}

\author{Jacobo Aguirre}
\email[]{jaguirre@cab.inta-csic.es}
\affiliation{Centro de Astrobiolog\'{\i}a (CSIC-INTA), ctra. de Ajalvir km 4, 28850 Torrej\'on de Ardoz, Madrid, Spain }
\affiliation{Grupo Interdisciplinar de Sistemas Complejos (GISC), Madrid, Spain}

 \begin{abstract}

{\bf Abstract:} Recent years have witnessed the detection of an increasing number of
complex organic molecules in interstellar space, some of them being of
prebiotic interest. Disentangling the origin of interstellar prebiotic
chemistry and its connection to biochemistry and ultimately to biology
is an enormously challenging scientific goal where the application of
complexity theory and network science has not been fully exploited.
Encouraged by this idea, we present a theoretical and computational
framework to model the evolution of simple networked structures toward
complexity. In our environment, complex networks represent simplified
chemical compounds, and interact optimizing the dynamical importance of
their nodes. We describe the emergence of a transition from simple
networks toward complexity when the parameter representing the
environment reaches a critical value. Notably, although our system
does not attempt to model the rules of real chemistry, nor is dependent
on external input data, the results describe the emergence of complexity
in the evolution of chemical diversity in the interstellar medium.
Furthermore, they reveal an as yet unknown relationship between the
abundances of molecules in dark clouds and the potential number of chemical
reactions that yield them as products, supporting the ability of the
conceptual framework presented here to shed light on real scenarios. Our
work reinforces the notion that some of the properties that condition
the extremely complex journey from the chemistry in space to prebiotic
chemistry and finally to life could show relatively simple and universal
patterns.\\

{\bf Significance statement:} {\it The road to life is punctuated by transitions toward complexity, 
from astrochemistry to biomolecules and eventually to living organisms. 
Disentangling the origin of such transitions is a challenge where the application 
of complexity and network theory has not been fully exploited. We introduce a computational 
framework in which simple networks simulate the most basic building bricks of life and interact to 
form complex structures, leading to an explosion of diversity when the parameter 
representing the environment reaches a critical value. While this model is abstract 
and unrelated to chemical theory, its predictions reliably mimic the molecular 
evolution in the interstellar medium during the transition toward chemical complexity, 
suggesting that the rules leading to the emergence of complexity may be universal.}
\end{abstract}

\maketitle
The origin of life on Earth is far away from being unveiled. Life could have appeared spontaneously in our early planet 
about 4 billion years ago,
or exist previously in the outer space and be brought by dust and meteoroids, as panspermia states.
An intermediate hypothesis is the
\textit{molecular panspermia}, which proposes that the original building blocks of life could have been produced in 
the interstellar medium (ISM) and be introduced in the early Earth by asteroids and meteorites during the Late Heavy Bombardment that took 
place between 3.8 and 4.1 billion years ago, importantly enriching prebiotic chemistry. 
Over 200 molecules have been detected in the ISM, with some of them being prebiotically relevant as e.g. glycolaldehyde, urea or ethanolamine 
\cite{Hollis2004,Belloche2019, Izaskun2020,Rivilla2021}. 
Prebiotic molecules such as glycine or ribose have indeed been found in meteorites and comets \cite{Altwegg2017,Furukawa2019}, 
which supports the idea that prebiotic species 
could initially form in interstellar space, and be transferred later on to planetesimals and 
to Earth during the formation of the Solar system. 

In parallel to the efforts to understand the origin of life from the biochemical optics, the development and advance of computation in
the last fifty years
propitiated the study of life modeled as a cellular automaton \cite{Neumann1966,Wolfram1983}. 
Despite the simplicity of the rules, the different systems under study, in particular Conway's Game of Life \cite{Conway1970},
presented an unexpected variety of spatio-temporal patterns and cast light on how complexity, 
emergence, and self-organization arise from a simple system. More complex systems were introduced in the 80's to model 
Darwinian evolution with the use of a new type of artificial 
life where organisms described as computer 
programs could self-replicate, adapt and mutate by natural selection, mostly competing for the control of the
memory of the computer (e.g. CoreWar \cite{rasmussen_knudsen_feldberg_hindsholm_1990} and Avida \cite{lenski_ofria_collier_adami_1999, Adami:2000}).
The introduction of these digital organisms to address fundamental biological questions was supported on two main statements.
First, they provide a way to generalize life beyond the organisms detected so-far in our biosphere.
Second, they allow to perform, enlarge and repeat experiments on a scale that is unachievable with real entities
\cite{lenski_ofria_collier_adami_1999}. 

In order to assess whether artificial 
life and its connection with complex networks theory can bring light to the study of the origin of life, in this work we present 
a computational framework, 
NetWorld, where networks interact following very simple local rules inspired in network science and game theory leading to a 
{\it chemistry of networks}. 
Our objective is to test using real astrochemical data whether a basic digital framework 
can reproduce certain general properties of the difficult transition from chemistry to biology, 
and therefore describe in an abstract level the creation of the basic building blocks of life.

\section{Results}

In NetWorld every chemical compound is represented by a network,
and this allows us to apply the strength and tools of 
complex network theory. 
From simply isolated nodes that simulate an initial state of total lack of complexity, 
this new artificial chemistry shows the emergence of a transition beyond which the environment permits the appearance of 
a rich variety of networks with different spectral, 
topological and dynamical properties, mimicking in a very simplified manner the first steps of prebiotic chemistry and its natural
evolution toward complexity. 
We will pay special attention to the descriptive and predictive ability of this framework, showing 
that the results throw light on the chemical evolution toward complexity of molecular clouds in the interstellar medium
and reveal a so-far-unknown relationship between the abundances of the molecules present 
in dark clouds and the number of chemical reactions that have them as products.

\begin{figure*}[!]
	\centering
	\includegraphics[scale=1]{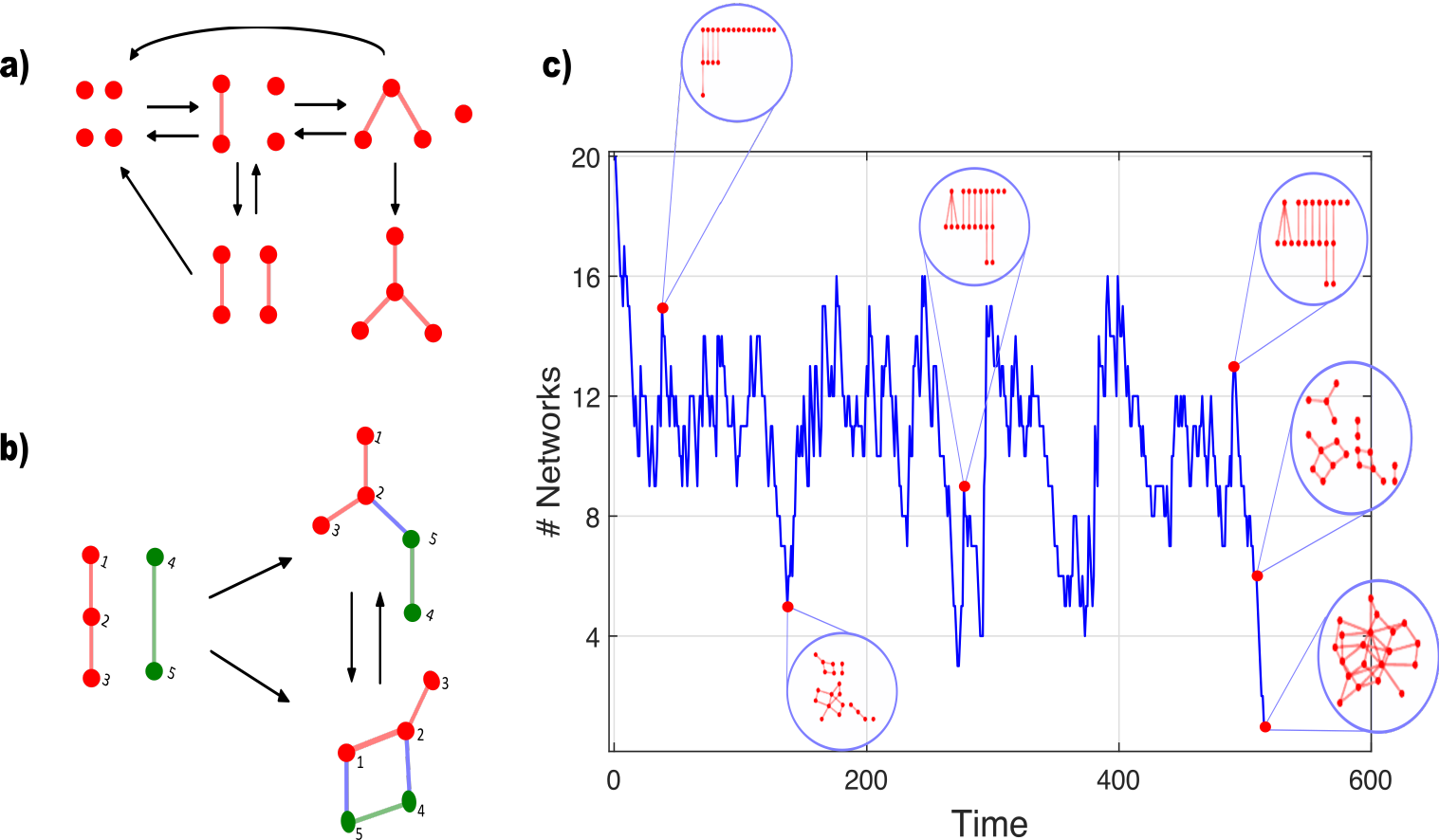}
	\caption{Description of the evolutionary dynamics created by NetWorld. 
	(a) Sketch of the potential evolution of a simple system formed by 4 nodes. 
	(b) In every time step of a simulation, two networks get in touch and interact until 
	the system reaches a final network (defined by a Nash equilibrium) or a cycle of several alternating structures. 
	The connector links (blue) join the original networks (red and green).
	(c) Example of a realization of the whole process with environment parameter $\beta=2.5$ and $N=20$ initial isolated nodes. 
	The process reaches an end when a single network of size 20 is created.}
	\label{fig1}
\end{figure*}

\subsection{Description and rules of the model}

An artificial chemistry model is defined by three components: a set 
of all possible structures, a set of rules that govern the interaction among structures and an 
algorithm that describes the reaction domain \cite{Dittrich2001}. Furthermore, depending on their level of abstraction, 
models can be classified in analogous, 
when they try to be faithful to natural chemistry, and abstract, otherwise. 
According to these definitions, NetWorld could be understood as
an extremely abstract artificial chemistry model, 
where nodes stand for indistinguishable basic entities --they are unweighted, and have no different 
properties to represent the chemical valency or size of atoms, for example--, and 
the bonds between nodes are represented by undirected and unweighted links. In a simple case these entities could be molecules where 
nodes stand for atoms and links for 
their interactions. However, as its internal rules are so different to those that govern real chemistry, 
our approach differs drastically from 
the large myriad of artificial chemistry models that quantitatively represent real processes through detailed descriptions
of the physico-chemical interaction between atoms \cite{Dittrich2001,Benko2003,Banzhaf2015}. 

Figure~\ref{fig1} presents a visual description of 
the evolutionary dynamics described by NetWorld, and Fig.~\ref{fig1}(a) shows a toy example of 4 initial nodes.
A rigorous explanation of the algorithm is presented in Supporting Information Texts S1 to S3, 
an analysis of the dependence of the computation time with NetWorld's parameters is introduced 
in Supporting Information Text S4 and information on 
the public availability of the code is found in Supporting Information Text S5.
Each process is started with an initial number $n(0)=N$ of isolated nodes. 
Nodes are neither created nor destroyed during the process. 
In each time step $t$, the population consists of $n(t)$ networks 
that will be made of the $N$ available nodes for the total ensemble. 
At the beginning of each time step $t$, two networks $A$ and $B$ of the total population are chosen randomly. 
They interact following a simple set of rules explained below and in 
Supporting Information Text S1.A. They either form a
new network $C$, simulating the reaction $A+B \rightarrow C$, or fail to join and remain as $A$ and $B$. In the latter case,
a different pair of networks $A'$ and $B'$
is chosen until one pair succeeds in forming a new network. At the end of the time step, every network $i$ in the population 
has a partition probability 
\begin{equation}
 P_i=\frac{2}{1+\exp(\mu_i \beta)}
 \label{Pi}
\end{equation}
of being divided into smaller pieces, simulating the reaction $C \rightarrow \sum D_j$ (see Supporting Information Text S1.B for a 
full description of the partition algorithm).
The stability parameter $\mu_i$ is 
the second smallest eigenvalue associated with the Laplacian matrix of network $i$. It is 
also known as the algebraic connectivity or the
Fiedler eigenvalue, 
and represents the resistance of a network to being split into different communities \cite{Newman:2010}. 
The environment parameter $\beta$ 
concentrates the whole physico-chemical properties of the environment, such as the temperature or the radiation.
$\beta$ is constant
during the whole process and it is the unique relevant 
global parameter of our model. When a new time step starts, the same process is repeated with the new collection of $n(t+1)$ networks. 
Note that $n(t+1)\geq n(t)-1$ and $n(t)\in[1,N]$ for all $t$. The process
finishes when (i) the totality of nodes collapse in a unique structure (i.e. $n(t)=1$), (ii) 
$n(t)>1$ but no networks will accept any new connections, or (iii) $t$ reaches a limit value of $10^4$ steps. A typical realization of the total process for an environment parameter 
$\beta=2.5$ and
$N=20$ initial nodes is plotted in Fig.~\ref{fig1}(c).

\begin{figure*}[!]
\centering
\includegraphics[scale=1]{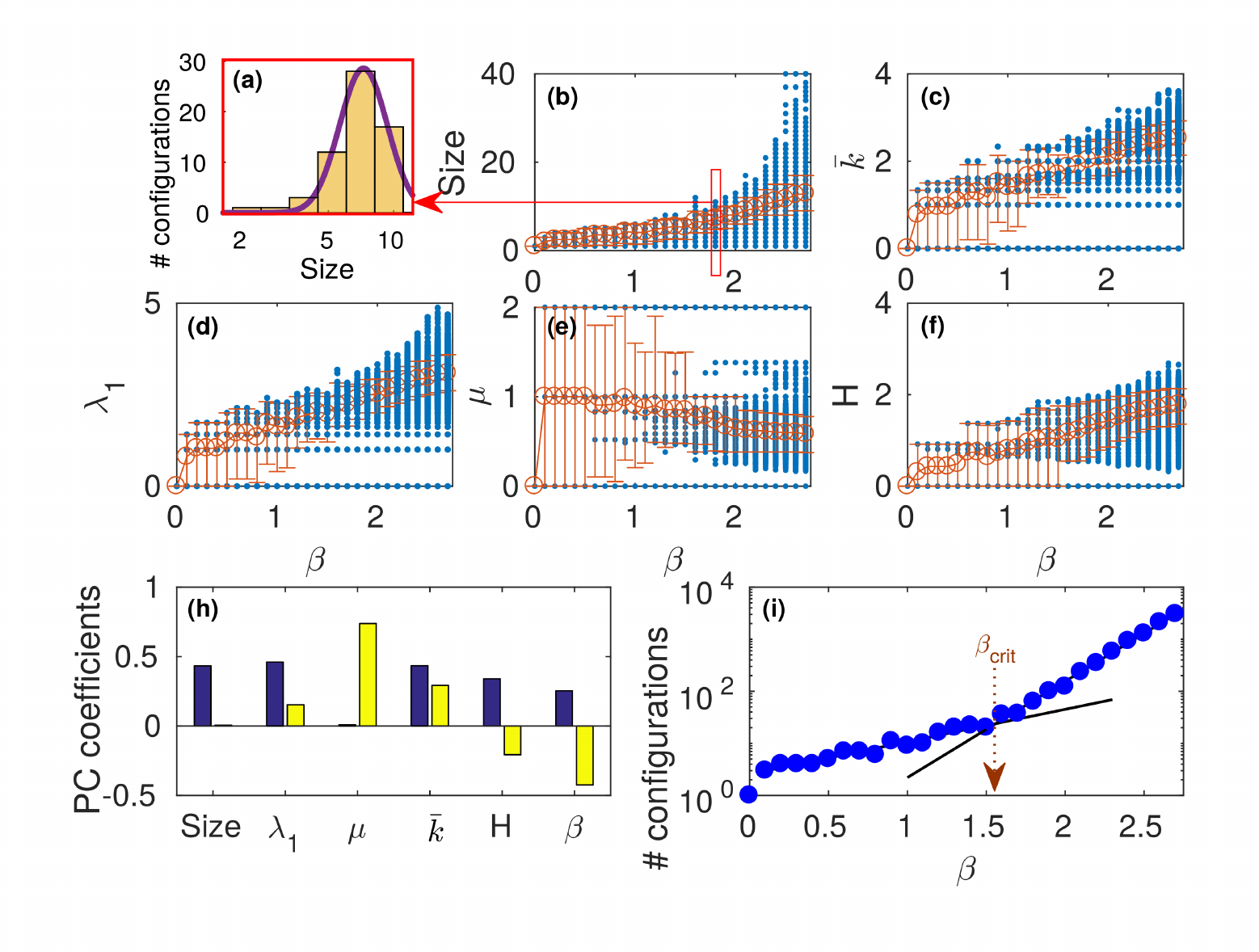} 
\caption{Description of the diversity of structures created by NetWorld for different values of the environment parameter
$\beta$ and $N=40$ initial nodes. 
(a) Histogram of network sizes (number of nodes in a network) for $\beta=1.8$. 
A log-normal fitting is plotted in bold line (goodness of 
fit of $r=0.993$). 
Dependence with $\beta$ of (b) size of the networks, (c) average degree $\vec{k}$, 
(d) maximum eigenvalue $\lambda_1$, (e) degree entropy $H$ (calculated as 
$H=-\sum_{i=1}^K p_i \times \log_2 (p_i)$, where $p_i$ is the fraction of nodes of degree $i$, 
and $K$ the maximum degree), and (f) stability parameter $\mu$. (g) First (blue, 54\% of variability) and second (yellow, 21\% of variability) components of
the PCA of the system. 
(h) Number of different configurations $N_{conf}$ as a function of $\beta$. Exponential approximations are plotted in dashed lines 
($N_{conf} = 2.56\exp(1.43\beta)\,, r=0.971$ for $\beta<\beta_{crit}$, and $N_{conf} = 0.03\exp(4.25\beta)\,,r=0.998$ 
for $\beta>\beta_{crit}$).
The critical value of $\beta$ ($\beta_{crit}\approx1.55$) is remarked. 
In (b-f) every single network is represented by a blue dot, 
and the average value and the 10-90 percentile range for each $\beta$ are plotted in red. 
}
	\label{fig2}
\end{figure*}

The interaction between the two networks $A$ and $B$ chosen randomly at time $t$ to give rise to network $C$ requires special attention
(see Fig.~\ref{fig1}(b)). 
It is inspired by a connecting method already applied to socio-economic networks \cite{Iranzo:2016,Iranzo2020},
and it is grounded in the extensive work devoted to
the description of the competition/cooperation between networks developed during the last decade 
for ordinary interactions \cite{Gomezgardenes:2012,Aguirre:2013,Wang:2014,Wang:2015}, 
or more recently for higher-order interactions \cite{Alvarez:2021,Kumar:2021,Bianconi:2021}.
We randomly choose one node $a$ in $A$ and one node $b$ in $B$ (connector nodes from now on) and connect them through an undirected
link (connector link from now on). This new link is accepted only if both connector nodes $a$ and $b$ 
increase their {\it dynamical importance} in the network, measured as 
$I= \lambda_1 u_l$.
$\lambda_1$ is the largest eigenvalue of the adjacency matrix of the new network formed by $A$, $B$ and the 
connector link just added, 
and $\vec{u}$ represents its associated eigenvector, L1-normalized such that $\sum u_k=1$.
$u_l$ is the eigenvector centrality of connector node $l$, and measures the importance of a node based on 
how well connected it is and how important its neighbors are. Both $\lambda_1$ and $\vec u$ are important
measurements of the dynamical properties associated with a complex network. If there was a dynamical process
of the type $\vec m(t+1)={\bf G} \vec m(t)$ evolving on a network of adjacency matrix {\bf G}, it is known that 
$\vec m(t)\rightarrow \vec u$ when $t\rightarrow\infty$,
and the population growth rate would be $\lambda_1$ \cite{Iranzo:2016}. 
The process of choosing
randomly a new pair of nodes $a'$ in $A$ and $b'$ in $B$ --already connected through a connector link from $a$ to $b$-- 
and checking whether they accept a link or not is repeated, taking into account that,
for simplicity and without loss of generality, only one connector link per node is allowed \cite{Iranzo2020}, and therefore
any pre-existing connector links associated with nodes $a'$ and $b'$ are erased. In general, this algorithm leads to the connection 
of $A$ and $B$ through a cascade and rewiring of connector links until the total network $C=A+B$ reaches a 
Nash equilibrium
or a cycle between several final configurations, and in the latter case one of the final cyclic configurations is chosen randomly as $C$.
If no links are accepted between $A$ and $B$ when all possible connections between both networks have been tried, we suppose that
they do not react and $A$ and $B$ remain unchanged.
In this case, the time is not increased and the interaction between two other networks starts.

In summary,
our framework is grounded in the simulation of abstract networked entities
that evolve following rules inherited from network science and game theory,
and there is a total absence of machine learning techniques or any kind of fitting with real data in the different steps of the model.

\subsection{Description of the artificial chemosphere created at NetWorld. Transition toward complexity}

Every simulation of the whole process throughout this work started from $N=40$ initial isolated nodes, 
lasted a maximum of $10^4$ time steps and was repeated 25 times for each value of the environment parameter $\beta$ in the range
$\beta=[0,2.7]$. The limit case $\beta\rightarrow \infty$ does not permit any network partition ($P=0$ in Eq.~\ref{Pi})
and is also studied.
All structures 
detected at the end of any time step for all realizations of each $\beta$, even if they were destroyed later, 
represent its diversity. 
The relative abundance of each configuration is given by the probability of finding it 
in the set of networks accumulated during 
all times and realizations (see Supporting Information Text S2 for a detailed explanation of how to 
compute the number of different configurations and the relative abundance in a simulation). 

\subsubsection{Diversity vs. environment}
We start focusing on how the environment shapes the networks emerging in the system. Figure~\ref{fig2}
shows the topological and structural description of the diversity created for
a wide range of values of the environment parameter. 
For $\beta=0$, every network of 2 nodes, i.e. the first to be created from isolated nodes,
immediately breaks up
and only isolated nodes are found for all times. 
When $\beta$ grows, 
the partition probability $P$ of a network decreases (Eq.~\ref{Pi}) and diverse structures emerge,
but it is not until large values of $\beta$ are reached ($\beta=2.5$ for $N=40$) that networks of the maximum possible size are created,
as shown in Fig.~\ref{fig2}(b).\footnote{Note that, in opposition to real chemistry, in our model all nodes are equal, 
and therefore there are very few potential configurations of small size (e.g. there is only 1 for size 2, while there is a large number 
of molecules of two atoms in nature).} The size, mean degree $\vec k$, 
largest eigenvalue of the adjacency matrix $\lambda_1$ and the degree entropy of the
created structures $H$, plotted in Figs.~\ref{fig2}(b-e), positively correlate with $\beta$, as shown 
in the first component of the Principal Components Analysis (PCA) calculated for the whole ensemble of existing 
structures and plotted in Fig.~\ref{fig2}(g) \cite{jolliffe_cadima_2016}. 
On the contrary, the average stability 
of the networks $\bar{\mu}$ correlates negatively with $\beta$
(Fig.~\ref{fig2}(f) and second component of the PCA in (Fig.~\ref{fig2}(g)), 
as more extensive and heterogeneous structures --i.e. with large entropy $H$-- are in general easier to divide, but the low partition probability
of networks for large $\beta$ permits their appearance and survival.

Importantly, in Fig.~\ref{fig2}(h) the growth of the number of different configurations with $\beta$ 
suffers a transition around $\beta_{crit}=1.55$ such that beyond that critical point the diversity explodes exponentially faster than before it. 
While the $10-90$ percentile range plotted in red in Figs.~\ref{fig2}(b-f) 
shows that the population is formed 
by a large number of networks with similar topological properties, for $\beta>\beta_{crit}$ there is also a {\it rare chemosphere} 
of structures that show very different topologies from the one of the majority, 
introducing a large amount of variability. 
Note that the existence of long tails of low-abundance entities is a typical property of real complex systems, 
notably ecological communities \cite{mcgill_2007,pedros2012}.
In summary, for low values of $\beta$ the main bricks of future complexity are formed, but only when
a critical state of the environment is reached may these motifs be sufficiently abundant and interact successfully to enrich the system 
with a large number of new structures. This new population is made of very diverse entities, where 
(i) regular/robust structures but also (ii) heterogeneous/less stable networks can de detected. 
The former are made out of very similar blocks that, 
from the optics of information theory, would not code complex information but 
show redundancy, a fundamental property to prevent attacks and failures when basic tasks must be developed. The latter
consist of complex networks where more information could be coded but are more sensitive to divisions, 
external perturbations, or losses of nodes.

\begin{figure}[t]
	\centering
	\includegraphics[width=\linewidth]{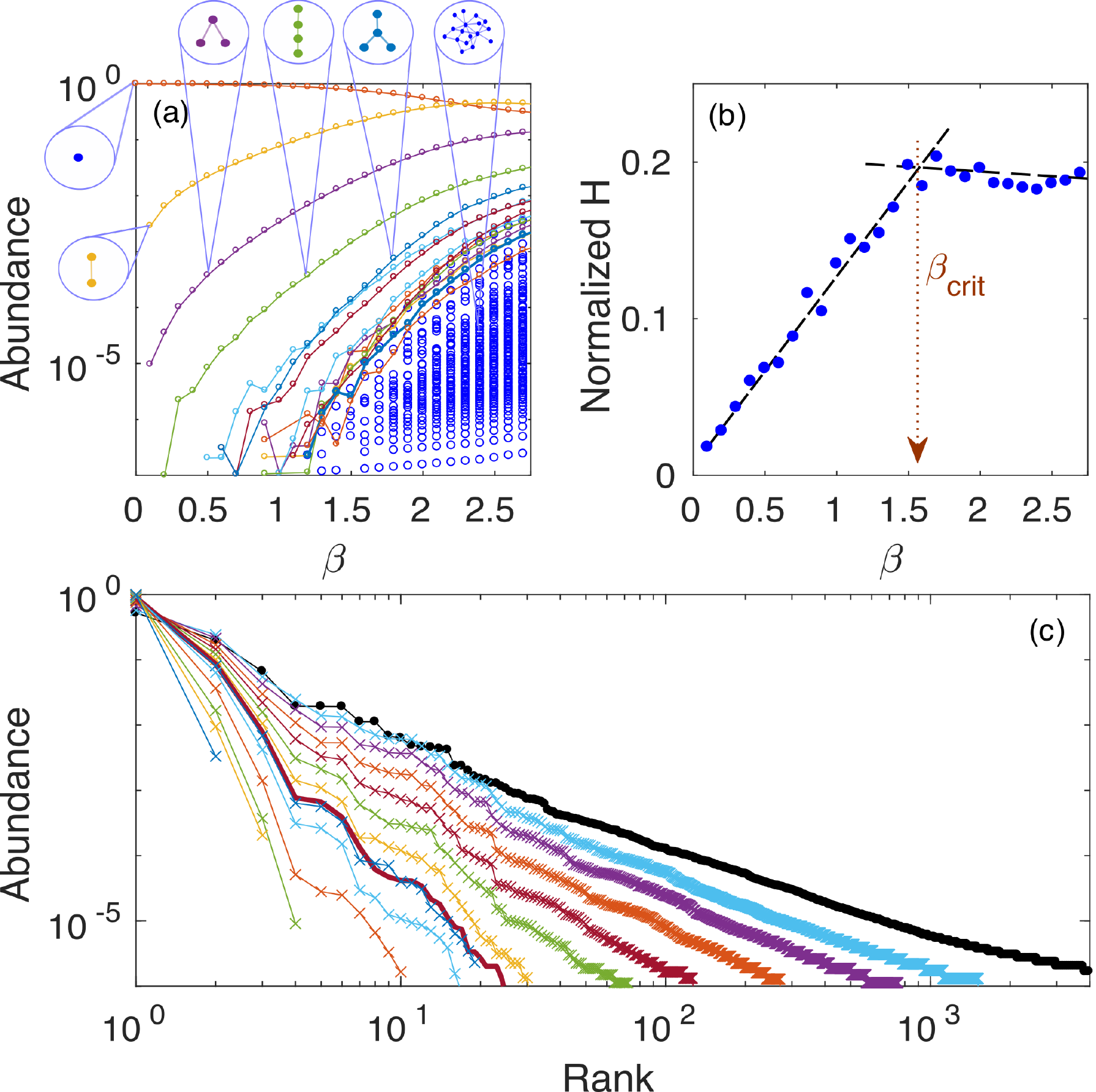} 
	\caption{Analysis of the relative abundance of structures created by NetWorld for different values of the environment parameter
	$\beta$ and $N=40$ initial nodes. (a) Abundance of each structure as a function of $\beta$.
	Each color represents a different structure, 
	but all structures emerged for 
	$\beta>\beta_{crit}$ are plotted in dark blue for clarity.
	(b) Normalized entropy $H_{norm}$ as a function of $\beta$. 
	$H_{norm}=\sum_{i=1}^{N_{conf}} p_i\times \log_2 p_i/\log_2 (N_{conf})$, where $p_i$ is the relative abundance of configuration $i$
and $N_{conf}$ is the number of different configurations. 
	Linear approximations are plotted in dashed lines:
	$H_{norm} = (0.006\pm0.004)+(0.121\pm0.005)\beta$ for $\beta<\beta_{crit}$, and $H_{norm} = (0.20\pm0.01)-(0.006\pm0.005)\beta$ 
	for $\beta>\beta_{crit}$.
	The critical value of $\beta_{crit}\approx1.55$ is remarked. 
	(c) Abundance rank for different values of $\beta$ (excluding isolated nodes). $\beta\in[0.1,2.7]$,
	$\beta$ grows from left to right in intervals of $\Delta \beta=0.2$. 
	The abundance rank for $\beta_{crit}=1.55$ is plotted in brown line, and that for $\beta\rightarrow \infty$ 
	(i.e., when the environment does not permit any network partition) is plotted in black circles.
		\label{fig3}}
\end{figure}

\subsubsection{Abundance vs. environment}

Figure \ref{fig3} describes the relative abundance of the networked structures obtained for the
different environments, that is, the probability of finding them 
in the set of networks accumulated during all times and realizations.
In Fig.~\ref{fig3}(a) we plot the relative abundance of each configuration 
as a function of $\beta$, and it is clear that the simplest structures are created at low values of 
$\beta$ and are especially abundant. A gradual appearance of larger and/or more complex configurations takes place for moderate $\beta$, 
and beyond $\beta=\beta_{crit}$ a cascade of new configurations 
leads to the exponential emergence of diversity already measured in Fig.~\ref{fig2}(h). 
The sharpness of the transition toward complexity
is specially visible in Fig.~\ref{fig3}(b), 
where the normalized entropy of the total ensemble of networks $H_{norm}$ \cite{gregori2016}
shows a transition
from a linear growth with $\beta$ to a constant value. 
For $\beta<\beta_{crit}$, the abundances of the few structures that exist
tend to be more uniformly distributed when $\beta$ grows (see Fig.~\ref{fig3}(a)), increasing the entropy of the system
(which reaches the maximum value $H_{norm}=1$ when all abundances are equal, see the mathematical expression for $H_{norm}$
in the caption of Fig.~\ref{fig3}).
However, once the critical environment is surpassed, 
the growth in complexity due to the tendency toward the relative abundance uniform distribution 
is balanced by the exponential emergence of diversity. 
Finally, in Fig.~\ref{fig3}(c) we plot the abundance ranks of the 
population for different
values of $\beta$. For low $\beta$, the curves are very skewed and show exponential decays, typical behavior of 
ecological environments with little diversity \cite{Fisher:2014}. When $\beta$ grows and crosses the critical value (brown line),
the curves gradually lose skewness and become power-laws, 
showing long tails consequence of a large diversity of rare configurations, and tend toward 
the limit case of the process in
which $\beta=\infty$ and the structures cannot divide.

\subsection{Application to a real scenario: Chemical complexity in the interstellar medium}

\begin{figure}[t]
	\centering
	\includegraphics[width=\linewidth]{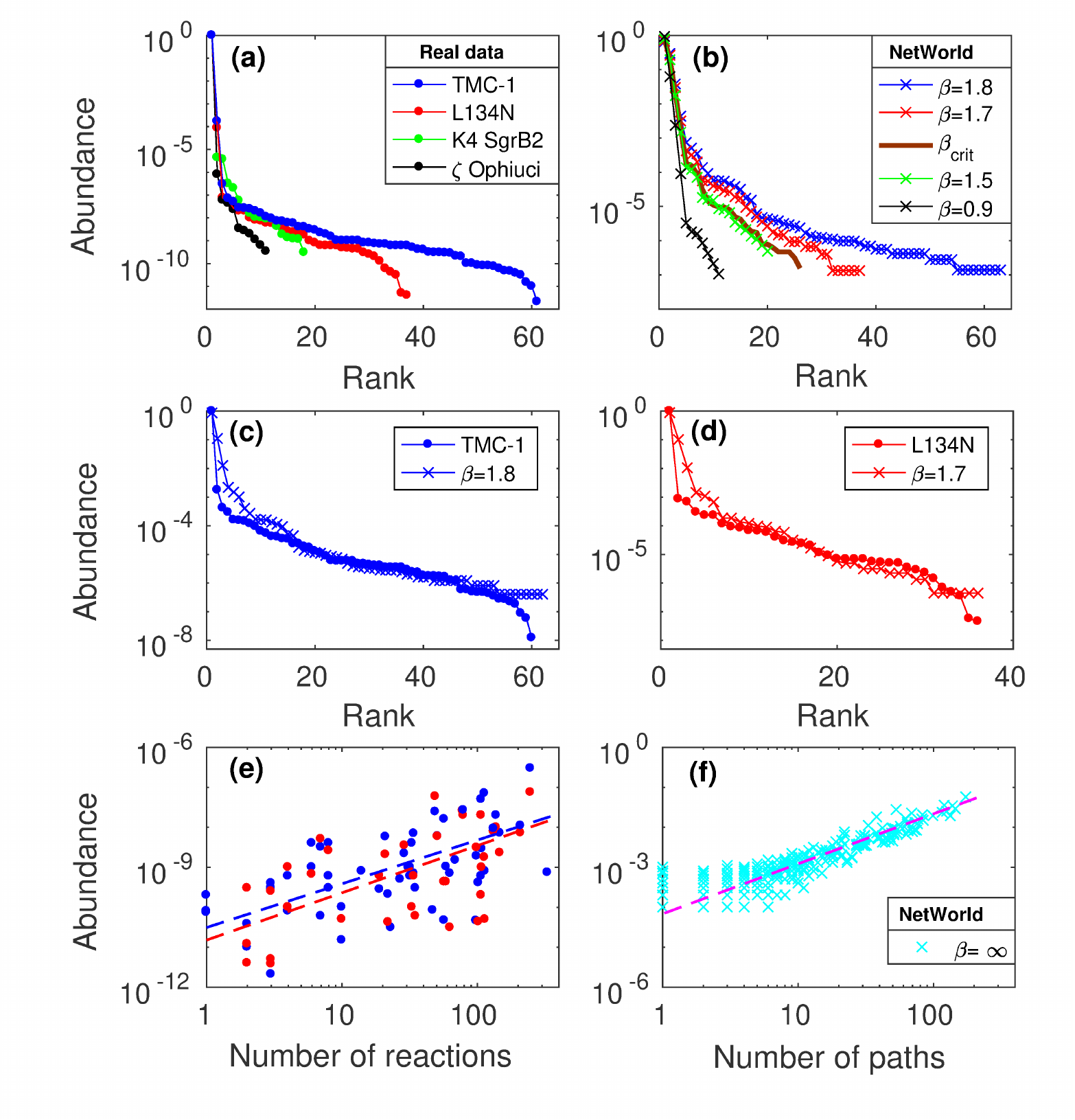}
	\caption{Comparison between the results of the digital environment NetWorld ($\times$)
	and the evolution of chemical complexity of interstellar clouds, the
	astrochemical environment where the most basic bricks of life are created ($\circ$).
	All abundance sets shown in this figure were L1-normalized (the sum is 1).
	(a) Abundance rank of the molecular compounds detected in four interstellar clouds: 
	the diffuse molecular cloud $\zeta$ Ophiuci \cite{Snow2006}, the translucent cloud located in the
	direction of the ultracompact HII region K4 in the SgrB2 molecular complex (a cloud in the transition 
	from the diffuse regime to the dense regime, see \cite{Thiel2017,Corby2018}), 
	and the dense clouds L134N (Serpens) and TMC-1 (Taurus) \cite{Agundez2013}.
	(b) Abundance rank obtained in NetWorld for $N=40$ initial nodes and environments 
	that are qualitatively compatible with the four sets of real data plotted in (a). 
	The abundance rank for $\beta_{crit}=1.55$ is plotted in brown line.
	(c) and (d) Comparison between the abundances in TMC-1 and L134N dense clouds with NetWorld simulations for $\beta=1.8$ and 1.7
	respectively. (e) Dependence of abundances of the molecules detected in L134N and TMC-1 
	on the number of astrophysical 
	reactions that have them as products. (f) Dependence of abundances of networks of size 10 created by NetWorld 
	(in the limit $\beta=\infty$
	for simplicity) on the number of paths used to create them.
	Curve fits are shown in (e) and (f) as dashed lines.
	Note that CO abundances are out of range in (e) for clarity, but were considered in the fits. 
		}
	\label{fig4}
\end{figure}
Molecules are an important component of the interstellar medium since they regulate its ionization state and energy dissipation. 
Molecules are typically found in interstellar clouds where the amount of interstellar dust, and thus of 
visual extinction A$_v$,
is large enough to prevent the photo-dissociation of molecular species by the external interstellar UV radiation field, which 
enables their formation and survival. The level of chemical complexity in interstellar clouds is however very different 
depending on their level of extinction, and on the available amount of molecular hydrogen (H$_2$) and carbon monoxide 
(CO) within them \cite{Snow2006}. In this way, interstellar clouds can be classified as diffuse atomic 
(with a fraction of H$_2$ with respect to total 
amount of atomic H of f(H$_2$)$<10\%$), diffuse molecular (f(H$_2$)$>10\%$), translucent (f(H$_2$)$>10\%$ and 
with a fraction of CO with respect to total amount of atomic C of f(CO)$<90\%$),
or dense clouds (f(H$_2$)$>10\%$ and f(CO)$>90\%$). The chemistry in diffuse atomic clouds is very limited \cite{Snow2006} 
while the chemistry in dense clouds presents a very high level of chemical complexity 
(see e.g. \cite{Cernicharo2021,McGuire2021}). Therefore,
we use here the molecular abundances measured toward diffuse molecular, translucent and dense clouds as test cases for the applicability 
of the digital environment NetWorld to real scenarios.

In Fig.~\ref{fig4}(a) we show the abundances of the chemical compounds detected toward four interstellar clouds ranked in order 
of their decreasing magnitude: 
(i) the interstellar cloud 
$\zeta$ Ophiuci, a diffuse molecular cloud where only a few molecules have been found \cite{Snow2006}; dust extinction is so weak 
($A_v=1.06$ mag)
that UV radiation
destroys most of the molecular material;
(ii) the translucent cloud located in the direction of the ultracompact HII region K4 in the SgrB2 
massive star-forming region \cite{Thiel2017, Corby2018}; 
this cloud has an extinction of A$_v$=2.0 mag, just enough to enable the formation of new molecular species and to 
protect the molecular content recently formed within the cloud, playing the role of the critical transition in the model; and 
(iii) L134N (Serpens) and (iv) TMC-1 (Taurus), two dense clouds
with $A_v>10$ mag
where an extensive number of both simple and complex molecules
have been synthesized 
thanks not only to the protection of the high A$_v$, but also to the high fraction of CO present \cite{Agundez2013}.
We refer to Supplementary Information Text S6 for the explanation on how we have obtained the molecular 
abundances toward the different clouds used in our analysis.
Fig.~\ref{fig4}(b) shows the abundance rank obtained with NetWorld
for 4 different environments that are qualitatively compatible with these four sets of real data: 
a low value of $\beta$ representing a harsh environment where most created compounds are rapidly destroyed, 
a value close to the critical $\beta_{crit}$ beyond which complexity expands, and two values of $\beta$ slightly over 
this transition point. We include the critical environment $\beta_{crit}=1.55$ for comparison.
In Figs.~\ref{fig4}(c) and (d) we focus on the potential quantitative agreement between real data and the results
of the artificial framework. We compare the abundance ranks for TMC-1 and L134N with those of NetWorld's $\beta=1.8$ and 1.7
respectively. The abundances are provided excluding the most frequent elements of each ensemble, H$_2$
in the molecular abundances and the isolated nodes in NetWorld.
The real and numerical curves show a quantitative agreement between
the number of molecules present in the cloud and the number of configurations in NetWorld, and also in the relative 
abundance for a large set of molecules and configurations.
The framework does not reproduce, however, the truncation shown in the real curves for the 2-3 lowest abundances, but
this behavior would disappear if new real data were introduced. Note that the astrochemical datasets 
of molecular species and their measured abundances
remain largely incomplete even for the most observed clouds such as TMC-1, 
and especially for low abundance species or very large molecules that are more difficult to detect 
(see e.g. \cite{Cernicharo2021,McGuire2018}).
It is also remarkable that the values of $\beta$ that best fit the astrochemical data of L134N and TMC-1 ($\beta=1.7$ and 1.8) are slightly 
beyond $\beta_{crit}=1.55$, and thus belong to a regime in which expansion of their chemical diversity is expected. 
Indeed, recent observational works toward the TMC-1 
dense cloud have revealed the presence of small polycyclic aromatic hydrocarbons (PAHs), demonstrating the ability of 
these environments to generate complex molecular structures \cite{Cernicharo2021,McGuire2018,McGuire2021}. 

Finally, a relevant pattern obtained in NetWorld is that the relative 
abundances of the different structures correlate with the number of paths 
identified to create them, following a very simple functional dependence of the type $y\propto x^\alpha$ 
(Fig.~\ref{fig4}(f), $\alpha=1.2\pm 0.2$
for networks of size 10, $N=40$ and $\beta=\infty$, see Supporting Information Text S3 for details on 
how to compute the number of paths to reach a configuration\footnote{We calculated the number of paths 
of each configuration for $\beta=\infty$ (with no loss of generality), as it is the case where there is no network partition 
and the computation is simpler.}). 
Note that this is a highly non-trivial result, as different paths have in general very different occurrence probabilities, 
and this expected diversity could in principle spoil the correlation.
In order to check whether a similar relationship might also emerge in astrochemical environments,
we plotted in Fig.~\ref{fig4}(e) the dependence of the molecular abundances measured toward the dense clouds L134N and TMC-1 
on the number of 
chemical reactions that have them as products, as a simple proxy for the
chemical paths. The number of chemical reactions is extracted from the astrochemical reaction dataset KIDA \cite{Wakelam2015}.
Surprisingly, 
the abundance data correlate with the number of reactions following the same functional dependence as NetWorld's simulations 
($y\propto x^\alpha$, where $\alpha_{TMC-1}=1.0 \pm0.2$, $r=0.57$, $p=2\cdot10^{-6}$ and $\alpha_{L134N}=1.2 
\pm 0.3$, $r=0.54$, 
$p=6\cdot10^{-4}$, see Supporting Information Text S7).
In addition, the $\alpha$-coefficient of the power-law that relates these two magnitudes is close to 1, 
indicating that the abundance of a certain molecule would be expected to be approximately proportional 
to the number of reactions that create it.

We caution, however, that the latter analysis could suffer from important biases. 
Small molecules that do not have dipole moment, such as N2 or CH4, cannot be observed at radio wavelengths. Furthermore, large molecules not only show low abundances but also their partition function is so large that it spreads the emission
across many energy levels, making the molecular line intensities very weak. 
Recent spectroscopic surveys carried out toward chemically rich sources such as SgrB2 N2 
\cite{Belloche2016}, IRAS16293-2422 \cite{Jorgensen2016}, TMC-1 
\cite{Cernicharo2021,McGuire2018} or G+0.693 
(see e.g. \cite{Izaskun2020, Rivilla2021}) with ALMA, 
the GBT or the Yebes 40m telescopes, have boosted the detection rate of 
new molecular species in the ISM in the past decade (see the review \cite{McGuire2022}), 
which is starting to alleviate this potential issue. Second, the astrochemical 
reaction databases are strongly biased to small molecules. 
All these chemical networks are based on the pioneering work of \cite{Prasad:1980}, 
whose goal was to investigate the interstellar chemistry of small molecular species. 
In addition, experimental and theoretical works of gas-phase reactions are extremely challenging
especially at the low temperatures typical of interstellar conditions, in particular of radical-radical 
reactions \cite{Shannon2013, Garcia2021}, and not all reactions yielding the same product are equally efficient due to the existence 
of, for example, high energy barriers. Because of all these reasons, the number of detected molecules 
and the datasets of reactions that can have them as products are inaccurate.

To address these limitations, we have statistically verified that the correlations 
between the molecular abundances and the number of reactions shown in Fig.~\ref{fig4}(e) 
for the interstellar clouds TMC-1 and L134N still hold when removing the effect of a controlling 
variable such as the molecular size, understood as the number of atoms contained within a
molecule (partial correlation coefficient $r'=0.49$ and $p'=7\cdot 10^{-5}$ for TMC-1 and $r'=0.42$ and $p'=0.007$ for L134N,
see Supporting Information Text S7 for more details). 
This result points to the correlation detected being independent of potential biases due to the molecular size.
A thorough statistical analysis, considering other astrochemical magnitudes 
related to the molecules found in different astrophysical environments, will be carried out in the future.

\section{Discussion} 

In this work we have introduced a conceptual and computational framework called NetWorld describing
the evolution of networked structures, where nodes interact following exclusively the optimization of their own dynamical importance.
Our results show that, although there is not a causal relationship between NetWorld’s framework and real astrochemical 
phenomena, a simple model grounded in network science and game theory captures key properties of the process 
toward chemical complexity and the creation of the building blocks of life. In particular, 
while our approach does not try to mimic real astrochemistry, 
it succeeds
to explain the emergence of interstellar molecular complexity, and points out as yet unknown astrochemical relationships 
that could be of importance in our understanding of the formation of prebiotic species in interstellar space.

Multiple methods have been proposed to map real chemistry to artificial chemistry models. 
Graphs, binary strings, character strings or numbers, among others, have been used to represent molecules 
\cite{farmer_autocatalytic_1986,banzhaf_self-replicating_1993,dittrich_self-evolution_1998,
dittrich_spontaneous_2000}.
Our networks describe molecules where atoms are nodes and 
their interactions are links. 
Beyond this mapping, one could add properties and impose more complex rules of interaction 
in order to get closer to real chemistry. Node labels and properties might be introduced to represent the atom type,
the hybridization type, charge, valency, or radicals. 
Moreover, an energy function could be used to choose the 
structure resulting from a reaction on the basis of the change of energy \cite{Benko2003}.
Based on the current state of our framework, adding these features in NetWorld is 
possible and, a priori, computationally feasible. 
However, we stress that fundamental features of NetWorld are its simplicity and abstraction, 
as we aim to present a framework that transcends chemistry to describe the interaction 
between complex structures of different nature and in diverse environments,
from nodes representing atoms to biomolecules or even species. 

In the same line of thought, the natural transition toward complexity emerged from our computational environment has also been observed in complex ecosystems. 
In particular, Fisher \& Mehta \cite{Fisher:2014} 
reported a strikingly similar pattern in which the skewness of biodiversity rank-abundance curves decreased 
and the overall diversity increased with carrying capacity, 
highlighting the sharp transition between stochastic, neutral regimes and selection-dominated, niche regimes. 
The environment parameter $\beta$ introduced here and the dust extinction A$_v$ 
in interstellar chemistry --two quantities directly related since the rate constants of interstellar 
UV photo-destruction reactions depend exponentially of $-$A$_v$ \cite{Holdship2017}--, 
could be understood as loose proxies for the carrying capacity in the ecological context, 
in the sense that a low $\beta$/low A$_v$ results in a ``harsher'' environment that limits network/molecules/species richness. 
Continuing the analogy, in the low $\beta$/low A$_v$ regime our network communities and the interstellar molecular abundances 
show a highly skewed abundance distribution 
(see Figs.~\ref{fig4}(a-b))
and are indeed dominated by stochasticity (the partition probability $P$ introduced in Eq.~\ref{Pi} 
and the interaction with UV radiation respectively). 
On the other hand, in the high $\beta$/high A$_v$
regime they present low skewness and high diversity, and are dominated by selection of structures with a higher number 
of paths/reactions leading to them (as shown in Figs.~\ref{fig4}(e-f)). 
All in all, we believe that (i) the similarities between the results in \cite{Fisher:2014}, based on models 
that are firmly rooted in classical ecological theory and checked with real data, 
(ii) those obtained from molecular abundances in interstellar clouds,
and (iii) the ones introduced by our computational environment, derived from a simple framework with no a priori ecological or chemical
assumptions, are not coincidental. They 
instead hint that the long path from the creation of the basic prebiotic compounds in the interstellar medium 
to the origin of life and its evolution on the early Earth 
could show universal patterns and common phenomenologies at all scales and across all stages.

Finally, while we have exclusively used sets of isolated and indistinguishable nodes as initial conditions, and have focused on
the description of the emerging diversity created, the framework
here introduced could be of use in many other contexts. Advancing in the subject of the origin and evolution of early life, 
potentially fruitful lines of future work could be the analysis of the interaction of small motifs 
to simulate the polymerization of simple chemical 
compounds, resembling the phenomenology present in the RNA World, or
the search for autocatalytic reactions that could help us advance toward a proto-replication of networks 
where the concepts of mutation and fitness
could be analysed as emerging properties of the system, instead of introducing them {\it ad-hoc} as has been done so far in the literature.

\begin{acknowledgments}
The authors acknowledge fruitful conversations with A. Aguirre-Tamaral, C. Briones, M. Castro, J. Garc\'{\i}a de la Concepci\'on,
J.A. Garc\'{\i}a-Mart\'{\i}n, R. Guantes, D. Hochberg, J. Iranzo, A. Luc\'{\i}a-Sanz, S. Manrubia and M. Ruiz-Bermejo, and technical support from N. Aguirre and J. Aguirre. 
I.J.-S. and J.A. received support from Projects PID2019-105552RB-C41, PID2021-122936NB-I00
and MDM-2017-0737 Unidad de Excelencia "María de Maeztu"-Centro de Astrobiología (CSIC-INTA)
by the Spanish Ministry of Science and Innovation/State Agency of Research MCIN/AEI/10.13039/501100011033 and
by "Fondo Europeo de Desarrollo Regional (FEDER) Una manera de hacer Europa", through Project ESP2017-86582-C4-1-R. 
F.P-S. was supported by the European Union’s Horizon 2020 research and innovation programme under the Marie Skłodowska-Curie grant agreement No 892961.
 \end{acknowledgments}

\subsection*{Author contributions}
M.G.-S., I.J.-S., F.P.-S., and J.A. designed the project and wrote the paper; M.G.-S., F.P.-S.,
and J.A. conceived the computational environment; M.G.-
S. and J.A. implemented the algorithm and performed the
numerical experiments; and I.J.-S. and J.A. contributed to
the applicability of the results to real astrochemical data.

\subsection*{Supporting Information}
The Supporting Information of this paper consists of Supporting Figs.~1-5 and Supporting Information Texts~S1-S7. 

It is accessible at \url{https://raw.githubusercontent.com/MiguelGarciaSanchez/NetWorld/main/garcia_sanchez_et_al_Supp_Info_2022.pdf}\\

The titles of Supporting Information Texts S1-S7 are:
\begin{itemize}
 \item S1. NetWorld's algorithm.
 \item S2. Computation of the diversity and the relative abundance of the different configurations for a fixed environment in NetWorld.
 \item S3. Computation of the number of paths to reach a configuration in NetWorld.
 \item S4. Dependence of the computation time with NetWorld's parameters.
 \item S5. Algorithm availability.
 \item S6. Molecular abundances measured toward diffuse molecular, translucent and dense clouds.
 \item S7. Statistical analysis of the data associated with molecular clouds TMC-1 and L134N.

\end{itemize}

\bibliography{garcia_sanchez_et_al}

\begin{thebibliography}{49}
\expandafter\ifx\csname natexlab\endcsname\relax\def\natexlab#1{#1}\fi
\expandafter\ifx\csname bibnamefont\endcsname\relax
  \def\bibnamefont#1{#1}\fi
\expandafter\ifx\csname bibfnamefont\endcsname\relax
  \def\bibfnamefont#1{#1}\fi
\expandafter\ifx\csname citenamefont\endcsname\relax
  \def\citenamefont#1{#1}\fi
\expandafter\ifx\csname url\endcsname\relax
  \def\url#1{\texttt{#1}}\fi
\expandafter\ifx\csname urlprefix\endcsname\relax\def\urlprefix{URL }\fi
\providecommand{\bibinfo}[2]{#2}
\providecommand{\eprint}[2][]{\url{#2}}

\bibitem[{\citenamefont{{Hollis} et~al.}(2004)\citenamefont{{Hollis}, {Jewell},
  {Lovas}, and {Remijan}}}]{Hollis2004}
\bibinfo{author}{\bibfnamefont{J.~M.} \bibnamefont{{Hollis}}},
  \bibinfo{author}{\bibfnamefont{P.~R.} \bibnamefont{{Jewell}}},
  \bibinfo{author}{\bibfnamefont{F.~J.} \bibnamefont{{Lovas}}},
  \bibnamefont{and}
  \bibinfo{author}{\bibfnamefont{A.}~\bibnamefont{{Remijan}}},
  \bibinfo{journal}{apjl} \textbf{\bibinfo{volume}{613}}, \bibinfo{pages}{L45}
  (\bibinfo{year}{2004}).

\bibitem[{\citenamefont{{Belloche, A.} et~al.}(2019)\citenamefont{{Belloche,
  A.}, {Garrod, R. T.}, {M\"uller, H. S. P.}, {Menten, K. M.}, {Medvedev, I.},
  {Thomas, J.}, and {Kisiel, Z.}}}]{Belloche2019}
\bibinfo{author}{\bibnamefont{{Belloche, A.}}},
  \bibinfo{author}{\bibnamefont{{Garrod, R. T.}}},
  \bibinfo{author}{\bibnamefont{{M\"uller, H. S. P.}}},
  \bibinfo{author}{\bibnamefont{{Menten, K. M.}}},
  \bibinfo{author}{\bibnamefont{{Medvedev, I.}}},
  \bibinfo{author}{\bibnamefont{{Thomas, J.}}}, \bibnamefont{and}
  \bibinfo{author}{\bibnamefont{{Kisiel, Z.}}}, \bibinfo{journal}{A\&A}
  \textbf{\bibinfo{volume}{628}}, \bibinfo{pages}{A10} (\bibinfo{year}{2019}),
  \urlprefix\url{https://doi.org/10.1051/0004-6361/201935428}.

\bibitem[{\citenamefont{Jiménez-Serra
  et~al.}(2020)\citenamefont{Jiménez-Serra, Martín-Pintado, Rivilla,
  Rodríguez-Almeida, Alonso~Alonso, Zeng, Cocinero, Martín, Requena-Torres,
  Martín-Domenech et~al.}}]{Izaskun2020}
\bibinfo{author}{\bibfnamefont{I.}~\bibnamefont{Jiménez-Serra}},
  \bibinfo{author}{\bibfnamefont{J.}~\bibnamefont{Martín-Pintado}},
  \bibinfo{author}{\bibfnamefont{V.~M.} \bibnamefont{Rivilla}},
  \bibinfo{author}{\bibfnamefont{L.}~\bibnamefont{Rodríguez-Almeida}},
  \bibinfo{author}{\bibfnamefont{E.~R.} \bibnamefont{Alonso~Alonso}},
  \bibinfo{author}{\bibfnamefont{S.}~\bibnamefont{Zeng}},
  \bibinfo{author}{\bibfnamefont{E.~J.} \bibnamefont{Cocinero}},
  \bibinfo{author}{\bibfnamefont{S.}~\bibnamefont{Martín}},
  \bibinfo{author}{\bibfnamefont{M.}~\bibnamefont{Requena-Torres}},
  \bibinfo{author}{\bibfnamefont{R.}~\bibnamefont{Martín-Domenech}},
  \bibnamefont{et~al.}, \bibinfo{journal}{Astrobiology}
  \textbf{\bibinfo{volume}{20}}, \bibinfo{pages}{1048} (\bibinfo{year}{2020}),
  \bibinfo{note}{pMID: 32283036},
  \eprint{https://doi.org/10.1089/ast.2019.2125},
  \urlprefix\url{https://doi.org/10.1089/ast.2019.2125}.

\bibitem[{\citenamefont{Rivilla et~al.}(2021)\citenamefont{Rivilla,
  Jim{\'e}nez-Serra, Mart{\'\i}n-Pintado, Briones, Rodr{\'\i}guez-Almeida,
  Rico-Villas, Tercero, Zeng, Colzi, de~Vicente et~al.}}]{Rivilla2021}
\bibinfo{author}{\bibfnamefont{V.~M.} \bibnamefont{Rivilla}},
  \bibinfo{author}{\bibfnamefont{I.}~\bibnamefont{Jim{\'e}nez-Serra}},
  \bibinfo{author}{\bibfnamefont{J.}~\bibnamefont{Mart{\'\i}n-Pintado}},
  \bibinfo{author}{\bibfnamefont{C.}~\bibnamefont{Briones}},
  \bibinfo{author}{\bibfnamefont{L.~F.} \bibnamefont{Rodr{\'\i}guez-Almeida}},
  \bibinfo{author}{\bibfnamefont{F.}~\bibnamefont{Rico-Villas}},
  \bibinfo{author}{\bibfnamefont{B.}~\bibnamefont{Tercero}},
  \bibinfo{author}{\bibfnamefont{S.}~\bibnamefont{Zeng}},
  \bibinfo{author}{\bibfnamefont{L.}~\bibnamefont{Colzi}},
  \bibinfo{author}{\bibfnamefont{P.}~\bibnamefont{de~Vicente}},
  \bibnamefont{et~al.}, \bibinfo{journal}{Proceedings of the National Academy
  of Sciences} \textbf{\bibinfo{volume}{118}} (\bibinfo{year}{2021}), ISSN
  \bibinfo{issn}{0027-8424},
  \eprint{https://www.pnas.org/content/118/22/e2101314118.full.pdf},
  \urlprefix\url{https://www.pnas.org/content/118/22/e2101314118}.

\bibitem[{\citenamefont{Altwegg et~al.}(2017)\citenamefont{Altwegg, Balsiger,
  Berthelier, Bieler, Calmonte, Fuselier, Goesmann, Gasc, Gombosi, Le~Roy
  et~al.}}]{Altwegg2017}
\bibinfo{author}{\bibfnamefont{K.}~\bibnamefont{Altwegg}},
  \bibinfo{author}{\bibfnamefont{H.}~\bibnamefont{Balsiger}},
  \bibinfo{author}{\bibfnamefont{J.}~\bibnamefont{Berthelier}},
  \bibinfo{author}{\bibfnamefont{A.}~\bibnamefont{Bieler}},
  \bibinfo{author}{\bibfnamefont{U.}~\bibnamefont{Calmonte}},
  \bibinfo{author}{\bibfnamefont{S.}~\bibnamefont{Fuselier}},
  \bibinfo{author}{\bibfnamefont{F.}~\bibnamefont{Goesmann}},
  \bibinfo{author}{\bibfnamefont{S.}~\bibnamefont{Gasc}},
  \bibinfo{author}{\bibfnamefont{T.~I.} \bibnamefont{Gombosi}},
  \bibinfo{author}{\bibfnamefont{L.}~\bibnamefont{Le~Roy}},
  \bibnamefont{et~al.}, \bibinfo{journal}{Monthly Notices of the Royal
  Astronomical Society} \textbf{\bibinfo{volume}{469}}, \bibinfo{pages}{S130}
  (\bibinfo{year}{2017}), ISSN \bibinfo{issn}{0035-8711},
  \eprint{https://academic.oup.com/mnras/article-pdf/469/Suppl\_2/S130/18563276/stx1415.pdf},
  \urlprefix\url{https://doi.org/10.1093/mnras/stx1415}.

\bibitem[{\citenamefont{Furukawa et~al.}(2019)\citenamefont{Furukawa,
  Chikaraishi, Ohkouchi, Ogawa, Glavin, Dworkin, Abe, and
  Nakamura}}]{Furukawa2019}
\bibinfo{author}{\bibfnamefont{Y.}~\bibnamefont{Furukawa}},
  \bibinfo{author}{\bibfnamefont{Y.}~\bibnamefont{Chikaraishi}},
  \bibinfo{author}{\bibfnamefont{N.}~\bibnamefont{Ohkouchi}},
  \bibinfo{author}{\bibfnamefont{N.~O.} \bibnamefont{Ogawa}},
  \bibinfo{author}{\bibfnamefont{D.~P.} \bibnamefont{Glavin}},
  \bibinfo{author}{\bibfnamefont{J.~P.} \bibnamefont{Dworkin}},
  \bibinfo{author}{\bibfnamefont{C.}~\bibnamefont{Abe}}, \bibnamefont{and}
  \bibinfo{author}{\bibfnamefont{T.}~\bibnamefont{Nakamura}},
  \bibinfo{journal}{Proceedings of the National Academy of Sciences}
  \textbf{\bibinfo{volume}{116}}, \bibinfo{pages}{24440}
  (\bibinfo{year}{2019}), ISSN \bibinfo{issn}{0027-8424},
  \eprint{https://www.pnas.org/content/116/49/24440.full.pdf},
  \urlprefix\url{https://www.pnas.org/content/116/49/24440}.

\bibitem[{\citenamefont{Neumann and Burks}(1966)}]{Neumann1966}
\bibinfo{author}{\bibfnamefont{J.~V.} \bibnamefont{Neumann}} \bibnamefont{and}
  \bibinfo{author}{\bibfnamefont{A.~W.} \bibnamefont{Burks}},
  \emph{\bibinfo{title}{Theory of Self-Reproducing Automata}}
  (\bibinfo{publisher}{University of Illinois Press}, \bibinfo{address}{USA},
  \bibinfo{year}{1966}).

\bibitem[{\citenamefont{Wolfram}(1983)}]{Wolfram1983}
\bibinfo{author}{\bibfnamefont{S.}~\bibnamefont{Wolfram}},
  \bibinfo{journal}{Rev. Mod. Phys.} \textbf{\bibinfo{volume}{55}},
  \bibinfo{pages}{601} (\bibinfo{year}{1983}),
  \urlprefix\url{https://link.aps.org/doi/10.1103/RevModPhys.55.601}.

\bibitem[{\citenamefont{Gardner}(1970)}]{Conway1970}
\bibinfo{author}{\bibfnamefont{M.}~\bibnamefont{Gardner}},
  \bibinfo{journal}{Scientific American} \textbf{\bibinfo{volume}{223}},
  \bibinfo{pages}{120} (\bibinfo{year}{1970}).

\bibitem[{\citenamefont{Rasmussen et~al.}(1990)\citenamefont{Rasmussen,
  Knudsen, Feldberg, and
  Hindsholm}}]{rasmussen_knudsen_feldberg_hindsholm_1990}
\bibinfo{author}{\bibfnamefont{S.}~\bibnamefont{Rasmussen}},
  \bibinfo{author}{\bibfnamefont{C.}~\bibnamefont{Knudsen}},
  \bibinfo{author}{\bibfnamefont{R.}~\bibnamefont{Feldberg}}, \bibnamefont{and}
  \bibinfo{author}{\bibfnamefont{M.}~\bibnamefont{Hindsholm}},
  \bibinfo{journal}{Physica D: Nonlinear Phenomena}
  \textbf{\bibinfo{volume}{42}}, \bibinfo{pages}{111–134}
  (\bibinfo{year}{1990}).

\bibitem[{\citenamefont{Lenski et~al.}(1999)\citenamefont{Lenski, Ofria,
  Collier, and Adami}}]{lenski_ofria_collier_adami_1999}
\bibinfo{author}{\bibfnamefont{R.~E.} \bibnamefont{Lenski}},
  \bibinfo{author}{\bibfnamefont{C.}~\bibnamefont{Ofria}},
  \bibinfo{author}{\bibfnamefont{T.~C.} \bibnamefont{Collier}},
  \bibnamefont{and} \bibinfo{author}{\bibfnamefont{C.}~\bibnamefont{Adami}},
  \bibinfo{journal}{Nature} \textbf{\bibinfo{volume}{400}},
  \bibinfo{pages}{661–664} (\bibinfo{year}{1999}).

\bibitem[{\citenamefont{Adami et~al.}(2000)\citenamefont{Adami, Ofria, and
  Collier}}]{Adami:2000}
\bibinfo{author}{\bibfnamefont{C.}~\bibnamefont{Adami}},
  \bibinfo{author}{\bibfnamefont{C.}~\bibnamefont{Ofria}}, \bibnamefont{and}
  \bibinfo{author}{\bibfnamefont{T.~C.} \bibnamefont{Collier}},
  \bibinfo{journal}{Proceedings of the National Academy of Sciences}
  \textbf{\bibinfo{volume}{97}}, \bibinfo{pages}{4463} (\bibinfo{year}{2000}),
  ISSN \bibinfo{issn}{0027-8424},
  \eprint{https://www.pnas.org/content/97/9/4463.full.pdf},
  \urlprefix\url{https://www.pnas.org/content/97/9/4463}.

\bibitem[{\citenamefont{Dittrich et~al.}(2001)\citenamefont{Dittrich, Ziegler,
  and Banzhaf}}]{Dittrich2001}
\bibinfo{author}{\bibfnamefont{P.}~\bibnamefont{Dittrich}},
  \bibinfo{author}{\bibfnamefont{J.}~\bibnamefont{Ziegler}}, \bibnamefont{and}
  \bibinfo{author}{\bibfnamefont{W.}~\bibnamefont{Banzhaf}},
  \bibinfo{journal}{Artificial Life} \textbf{\bibinfo{volume}{7}},
  \bibinfo{pages}{225} (\bibinfo{year}{2001}).

\bibitem[{\citenamefont{Benkö et~al.}(2003)\citenamefont{Benkö, Flamm, and
  Stadler}}]{Benko2003}
\bibinfo{author}{\bibfnamefont{G.}~\bibnamefont{Benkö}},
  \bibinfo{author}{\bibfnamefont{C.}~\bibnamefont{Flamm}}, \bibnamefont{and}
  \bibinfo{author}{\bibfnamefont{P.~F.} \bibnamefont{Stadler}},
  \bibinfo{journal}{Journal of Chemical Information and Computer Sciences}
  \textbf{\bibinfo{volume}{43}}, \bibinfo{pages}{1085} (\bibinfo{year}{2003}).

\bibitem[{\citenamefont{Banzhaf and Yamamoto}(2010)}]{Banzhaf2015}
\bibinfo{author}{\bibfnamefont{W.}~\bibnamefont{Banzhaf}} \bibnamefont{and}
  \bibinfo{author}{\bibfnamefont{L.}~\bibnamefont{Yamamoto}},
  \emph{\bibinfo{title}{Artificial Chemistries}} (\bibinfo{publisher}{The MIT
  Press}, \bibinfo{address}{Cambridge, MA, USA}, \bibinfo{year}{2010}).

\bibitem[{\citenamefont{Newman}(2010)}]{Newman:2010}
\bibinfo{author}{\bibfnamefont{M.~E.~J.} \bibnamefont{Newman}},
  \emph{\bibinfo{title}{Networks: An Introduction}} (\bibinfo{publisher}{Oxford
  University Press, Inc.}, \bibinfo{address}{New York, NY, USA},
  \bibinfo{year}{2010}).

\bibitem[{\citenamefont{Iranzo et~al.}(2016)\citenamefont{Iranzo, Buld{\'u},
  and Aguirre}}]{Iranzo:2016}
\bibinfo{author}{\bibfnamefont{J.}~\bibnamefont{Iranzo}},
  \bibinfo{author}{\bibfnamefont{J.~M.} \bibnamefont{Buld{\'u}}},
  \bibnamefont{and} \bibinfo{author}{\bibfnamefont{J.}~\bibnamefont{Aguirre}},
  \bibinfo{journal}{Nature Communications} \textbf{\bibinfo{volume}{7}},
  \bibinfo{pages}{13273} (\bibinfo{year}{2016}).

\bibitem[{\citenamefont{Iranzo et~al.}(2020)\citenamefont{Iranzo,
  Pablo-Mart\'{\i}, and Aguirre}}]{Iranzo2020}
\bibinfo{author}{\bibfnamefont{J.}~\bibnamefont{Iranzo}},
  \bibinfo{author}{\bibfnamefont{F.}~\bibnamefont{Pablo-Mart\'{\i}}},
  \bibnamefont{and} \bibinfo{author}{\bibfnamefont{J.}~\bibnamefont{Aguirre}},
  \bibinfo{journal}{Phys. Rev. Research} \textbf{\bibinfo{volume}{2}},
  \bibinfo{pages}{043352} (\bibinfo{year}{2020}),
  \urlprefix\url{https://link.aps.org/doi/10.1103/PhysRevResearch.2.043352}.

\bibitem[{\citenamefont{G{\'o}mez-Garde{\~n}es
  et~al.}(2012)\citenamefont{G{\'o}mez-Garde{\~n}es, Reinares, Arenas, and
  Floria}}]{Gomezgardenes:2012}
\bibinfo{author}{\bibfnamefont{J.}~\bibnamefont{G{\'o}mez-Garde{\~n}es}},
  \bibinfo{author}{\bibfnamefont{I.}~\bibnamefont{Reinares}},
  \bibinfo{author}{\bibfnamefont{A.}~\bibnamefont{Arenas}}, \bibnamefont{and}
  \bibinfo{author}{\bibfnamefont{L.~M.} \bibnamefont{Floria}},
  \bibinfo{journal}{Sci Rep} \textbf{\bibinfo{volume}{2}}, \bibinfo{pages}{620}
  (\bibinfo{year}{2012}).

\bibitem[{\citenamefont{Aguirre et~al.}(2013)\citenamefont{Aguirre, Papo, and
  Buld{\'u}}}]{Aguirre:2013}
\bibinfo{author}{\bibfnamefont{J.}~\bibnamefont{Aguirre}},
  \bibinfo{author}{\bibfnamefont{D.}~\bibnamefont{Papo}}, \bibnamefont{and}
  \bibinfo{author}{\bibfnamefont{J.~M.} \bibnamefont{Buld{\'u}}},
  \bibinfo{journal}{Nature Physics} \textbf{\bibinfo{volume}{9}},
  \bibinfo{pages}{230} (\bibinfo{year}{2013}).

\bibitem[{\citenamefont{Wang et~al.}(2014)\citenamefont{Wang, Szolnoki, and
  Perc}}]{Wang:2014}
\bibinfo{author}{\bibfnamefont{Z.}~\bibnamefont{Wang}},
  \bibinfo{author}{\bibfnamefont{A.}~\bibnamefont{Szolnoki}}, \bibnamefont{and}
  \bibinfo{author}{\bibfnamefont{M.}~\bibnamefont{Perc}}, \bibinfo{journal}{New
  Journal of Physics} \textbf{\bibinfo{volume}{16}}, \bibinfo{pages}{033041}
  (\bibinfo{year}{2014}).

\bibitem[{\citenamefont{Wang et~al.}(2015)\citenamefont{Wang, Wang, Szolnoki,
  and Perc}}]{Wang:2015}
\bibinfo{author}{\bibfnamefont{Z.}~\bibnamefont{Wang}},
  \bibinfo{author}{\bibfnamefont{L.}~\bibnamefont{Wang}},
  \bibinfo{author}{\bibfnamefont{A.}~\bibnamefont{Szolnoki}}, \bibnamefont{and}
  \bibinfo{author}{\bibfnamefont{M.}~\bibnamefont{Perc}}, \bibinfo{journal}{The
  European Physical Journal B} \textbf{\bibinfo{volume}{88}},
  \bibinfo{pages}{124} (\bibinfo{year}{2015}).

\bibitem[{\citenamefont{Alvarez-Rodriguez
  et~al.}(2021)\citenamefont{Alvarez-Rodriguez, Battiston, de~Arruda, Moreno,
  Perc, and Latora}}]{Alvarez:2021}
\bibinfo{author}{\bibfnamefont{U.}~\bibnamefont{Alvarez-Rodriguez}},
  \bibinfo{author}{\bibfnamefont{F.}~\bibnamefont{Battiston}},
  \bibinfo{author}{\bibfnamefont{G.}~\bibnamefont{de~Arruda}},
  \bibinfo{author}{\bibfnamefont{Y.}~\bibnamefont{Moreno}},
  \bibinfo{author}{\bibfnamefont{M.}~\bibnamefont{Perc}}, \bibnamefont{and}
  \bibinfo{author}{\bibfnamefont{V.}~\bibnamefont{Latora}},
  \bibinfo{journal}{Nature Human Behavior} \textbf{\bibinfo{volume}{5}},
  \bibinfo{pages}{586} (\bibinfo{year}{2021}).

\bibitem[{\citenamefont{Kumar et~al.}(2021)\citenamefont{Kumar, Chowdhary,
  Capraro, and Perc}}]{Kumar:2021}
\bibinfo{author}{\bibfnamefont{A.}~\bibnamefont{Kumar}},
  \bibinfo{author}{\bibfnamefont{S.}~\bibnamefont{Chowdhary}},
  \bibinfo{author}{\bibfnamefont{V.}~\bibnamefont{Capraro}}, \bibnamefont{and}
  \bibinfo{author}{\bibfnamefont{M.~c.~v.} \bibnamefont{Perc}},
  \bibinfo{journal}{Phys. Rev. E} \textbf{\bibinfo{volume}{104}},
  \bibinfo{pages}{054308} (\bibinfo{year}{2021}),
  \urlprefix\url{https://link.aps.org/doi/10.1103/PhysRevE.104.054308}.

\bibitem[{\citenamefont{Bianconi}(2021)}]{Bianconi:2021}
\bibinfo{author}{\bibfnamefont{G.}~\bibnamefont{Bianconi}},
  \emph{\bibinfo{title}{Higher-Order Networks}} (\bibinfo{publisher}{Cambridge
  University Press}, \bibinfo{year}{2021}).

\bibitem[{\citenamefont{Jolliffe and Cadima}(2016)}]{jolliffe_cadima_2016}
\bibinfo{author}{\bibfnamefont{I.~T.} \bibnamefont{Jolliffe}} \bibnamefont{and}
  \bibinfo{author}{\bibfnamefont{J.}~\bibnamefont{Cadima}},
  \bibinfo{journal}{Philosophical Transactions of the Royal Society A:
  Mathematical, Physical and Engineering Sciences}
  \textbf{\bibinfo{volume}{374}}, \bibinfo{pages}{20150202}
  (\bibinfo{year}{2016}).

\bibitem[{\citenamefont{Mcgill et~al.}(2007)\citenamefont{Mcgill, Etienne,
  Gray, Alonso, Anderson, Benecha, Dornelas, Enquist, Green, He
  et~al.}}]{mcgill_2007}
\bibinfo{author}{\bibfnamefont{B.~J.} \bibnamefont{Mcgill}},
  \bibinfo{author}{\bibfnamefont{R.~S.} \bibnamefont{Etienne}},
  \bibinfo{author}{\bibfnamefont{J.~S.} \bibnamefont{Gray}},
  \bibinfo{author}{\bibfnamefont{D.}~\bibnamefont{Alonso}},
  \bibinfo{author}{\bibfnamefont{M.~J.} \bibnamefont{Anderson}},
  \bibinfo{author}{\bibfnamefont{H.~K.} \bibnamefont{Benecha}},
  \bibinfo{author}{\bibfnamefont{M.}~\bibnamefont{Dornelas}},
  \bibinfo{author}{\bibfnamefont{B.~J.} \bibnamefont{Enquist}},
  \bibinfo{author}{\bibfnamefont{J.~L.} \bibnamefont{Green}},
  \bibinfo{author}{\bibfnamefont{F.}~\bibnamefont{He}}, \bibnamefont{et~al.},
  \bibinfo{journal}{Ecology Letters} \textbf{\bibinfo{volume}{10}},
  \bibinfo{pages}{995–1015} (\bibinfo{year}{2007}).

\bibitem[{\citenamefont{Pedr\'os-Ali\'o}(2012)}]{pedros2012}
\bibinfo{author}{\bibfnamefont{C.}~\bibnamefont{Pedr\'os-Ali\'o}},
  \bibinfo{journal}{Annual Review of Marine Science}
  \textbf{\bibinfo{volume}{4}}, \bibinfo{pages}{449} (\bibinfo{year}{2012}),
  \bibinfo{note}{pMID: 22457983},
  \eprint{https://doi.org/10.1146/annurev-marine-120710-100948},
  \urlprefix\url{https://doi.org/10.1146/annurev-marine-120710-100948}.

\bibitem[{\citenamefont{Gregori et~al.}(2016)\citenamefont{Gregori, Perales,
  Rodriguez-Frias, Esteban, Quer, and Domingo}}]{gregori2016}
\bibinfo{author}{\bibfnamefont{J.}~\bibnamefont{Gregori}},
  \bibinfo{author}{\bibfnamefont{C.}~\bibnamefont{Perales}},
  \bibinfo{author}{\bibfnamefont{F.}~\bibnamefont{Rodriguez-Frias}},
  \bibinfo{author}{\bibfnamefont{J.~I.} \bibnamefont{Esteban}},
  \bibinfo{author}{\bibfnamefont{J.}~\bibnamefont{Quer}}, \bibnamefont{and}
  \bibinfo{author}{\bibfnamefont{E.}~\bibnamefont{Domingo}},
  \bibinfo{journal}{Virology} \textbf{\bibinfo{volume}{493}},
  \bibinfo{pages}{227} (\bibinfo{year}{2016}), ISSN \bibinfo{issn}{0042-6822},
  \urlprefix\url{https://www.sciencedirect.com/science/article/pii/S004268221630037X}.

\bibitem[{\citenamefont{Fisher and Mehta}(2014)}]{Fisher:2014}
\bibinfo{author}{\bibfnamefont{C.~K.} \bibnamefont{Fisher}} \bibnamefont{and}
  \bibinfo{author}{\bibfnamefont{P.}~\bibnamefont{Mehta}},
  \bibinfo{journal}{Proceedings of the National Academy of Sciences}
  \textbf{\bibinfo{volume}{111}}, \bibinfo{pages}{13111}
  (\bibinfo{year}{2014}), ISSN \bibinfo{issn}{0027-8424},
  \eprint{https://www.pnas.org/content/111/36/13111.full.pdf},
  \urlprefix\url{https://www.pnas.org/content/111/36/13111}.

\bibitem[{\citenamefont{Snow and McCall}(2006)}]{Snow2006}
\bibinfo{author}{\bibfnamefont{T.~P.} \bibnamefont{Snow}} \bibnamefont{and}
  \bibinfo{author}{\bibfnamefont{B.~J.} \bibnamefont{McCall}},
  \bibinfo{journal}{Annual Review of Astronomy and Astrophysics}
  \textbf{\bibinfo{volume}{44}}, \bibinfo{pages}{367} (\bibinfo{year}{2006}).

\bibitem[{\citenamefont{{Thiel, V.} et~al.}(2017)\citenamefont{{Thiel, V.},
  {Belloche, A.}, {Menten, K. M.}, {Garrod, R. T.}, and {M\"uller, H. S.
  P.}}}]{Thiel2017}
\bibinfo{author}{\bibnamefont{{Thiel, V.}}},
  \bibinfo{author}{\bibnamefont{{Belloche, A.}}},
  \bibinfo{author}{\bibnamefont{{Menten, K. M.}}},
  \bibinfo{author}{\bibnamefont{{Garrod, R. T.}}}, \bibnamefont{and}
  \bibinfo{author}{\bibnamefont{{M\"uller, H. S. P.}}}, \bibinfo{journal}{A\&A}
  \textbf{\bibinfo{volume}{605}}, \bibinfo{pages}{L6} (\bibinfo{year}{2017}),
  \urlprefix\url{https://doi.org/10.1051/0004-6361/201731495}.

\bibitem[{\citenamefont{{Corby, J. F.} et~al.}(2018)\citenamefont{{Corby, J.
  F.}, {McGuire, B. A.}, {Herbst, E.}, and {Remijan, A. J.}}}]{Corby2018}
\bibinfo{author}{\bibnamefont{{Corby, J. F.}}},
  \bibinfo{author}{\bibnamefont{{McGuire, B. A.}}},
  \bibinfo{author}{\bibnamefont{{Herbst, E.}}}, \bibnamefont{and}
  \bibinfo{author}{\bibnamefont{{Remijan, A. J.}}}, \bibinfo{journal}{Astronomy
  \& Astrophysics} \textbf{\bibinfo{volume}{610}}, \bibinfo{pages}{A10}
  (\bibinfo{year}{2018}),
  \urlprefix\url{https://doi.org/10.1051/0004-6361/201730988}.

\bibitem[{\citenamefont{Ag\'undez and Wakelam}(2013)}]{Agundez2013}
\bibinfo{author}{\bibfnamefont{M.}~\bibnamefont{Ag\'undez}} \bibnamefont{and}
  \bibinfo{author}{\bibfnamefont{V.}~\bibnamefont{Wakelam}},
  \bibinfo{journal}{Chem. Rev.} \textbf{\bibinfo{volume}{113}},
  \bibinfo{pages}{8710} (\bibinfo{year}{2013}).

\bibitem[{\citenamefont{{Cernicharo, J.}
  et~al.}(2021)\citenamefont{{Cernicharo, J.}, {Ag\'undez, M.}, {Cabezas, C.},
  {Tercero, B.}, {Marcelino, N.}, {Pardo, J. R.}, and {de Vicente,
  P.}}}]{Cernicharo2021}
\bibinfo{author}{\bibnamefont{{Cernicharo, J.}}},
  \bibinfo{author}{\bibnamefont{{Ag\'undez, M.}}},
  \bibinfo{author}{\bibnamefont{{Cabezas, C.}}},
  \bibinfo{author}{\bibnamefont{{Tercero, B.}}},
  \bibinfo{author}{\bibnamefont{{Marcelino, N.}}},
  \bibinfo{author}{\bibnamefont{{Pardo, J. R.}}}, \bibnamefont{and}
  \bibinfo{author}{\bibnamefont{{de Vicente, P.}}}, \bibinfo{journal}{A\&A}
  \textbf{\bibinfo{volume}{649}}, \bibinfo{pages}{L15} (\bibinfo{year}{2021}),
  \urlprefix\url{https://doi.org/10.1051/0004-6361/202141156}.

\bibitem[{\citenamefont{McGuire et~al.}(2021)\citenamefont{McGuire, Loomis,
  Burkhardt, Lee, Shingledecker, Charnley, Cooke, Cordiner, Herbst, Kalenskii
  et~al.}}]{McGuire2021}
\bibinfo{author}{\bibfnamefont{B.~A.} \bibnamefont{McGuire}},
  \bibinfo{author}{\bibfnamefont{R.~A.} \bibnamefont{Loomis}},
  \bibinfo{author}{\bibfnamefont{A.~M.} \bibnamefont{Burkhardt}},
  \bibinfo{author}{\bibfnamefont{K.~L.~K.} \bibnamefont{Lee}},
  \bibinfo{author}{\bibfnamefont{C.~N.} \bibnamefont{Shingledecker}},
  \bibinfo{author}{\bibfnamefont{S.~B.} \bibnamefont{Charnley}},
  \bibinfo{author}{\bibfnamefont{I.~R.} \bibnamefont{Cooke}},
  \bibinfo{author}{\bibfnamefont{M.~A.} \bibnamefont{Cordiner}},
  \bibinfo{author}{\bibfnamefont{E.}~\bibnamefont{Herbst}},
  \bibinfo{author}{\bibfnamefont{S.}~\bibnamefont{Kalenskii}},
  \bibnamefont{et~al.}, \bibinfo{journal}{Science}
  \textbf{\bibinfo{volume}{371}}, \bibinfo{pages}{1265} (\bibinfo{year}{2021}).

\bibitem[{\citenamefont{McGuire}(2018)}]{McGuire2018}
\bibinfo{author}{\bibfnamefont{B.~A.} \bibnamefont{McGuire}},
  \textbf{\bibinfo{volume}{239}}, \bibinfo{pages}{17} (\bibinfo{year}{2018}),
  \urlprefix\url{https://doi.org/10.3847/1538-4365/aae5d2}.

\bibitem[{\citenamefont{Wakelam et~al.}(2015)\citenamefont{Wakelam, Loison,
  Herbst, Pavone, Bergeat, B{\'{e}}roff, Chabot, Faure, Galli, Geppert
  et~al.}}]{Wakelam2015}
\bibinfo{author}{\bibfnamefont{V.}~\bibnamefont{Wakelam}},
  \bibinfo{author}{\bibfnamefont{J.-C.} \bibnamefont{Loison}},
  \bibinfo{author}{\bibfnamefont{E.}~\bibnamefont{Herbst}},
  \bibinfo{author}{\bibfnamefont{B.}~\bibnamefont{Pavone}},
  \bibinfo{author}{\bibfnamefont{A.}~\bibnamefont{Bergeat}},
  \bibinfo{author}{\bibfnamefont{K.}~\bibnamefont{B{\'{e}}roff}},
  \bibinfo{author}{\bibfnamefont{M.}~\bibnamefont{Chabot}},
  \bibinfo{author}{\bibfnamefont{A.}~\bibnamefont{Faure}},
  \bibinfo{author}{\bibfnamefont{D.}~\bibnamefont{Galli}},
  \bibinfo{author}{\bibfnamefont{W.~D.} \bibnamefont{Geppert}},
  \bibnamefont{et~al.}, \bibinfo{journal}{ApJS} \textbf{\bibinfo{volume}{217}},
  \bibinfo{pages}{20} (\bibinfo{year}{2015}),
  \urlprefix\url{https://doi.org/10.1088/0067-0049/217/2/20}.

\bibitem[{\citenamefont{{Belloche, A.} et~al.}(2016)\citenamefont{{Belloche,
  A.}, {M\"uller, H. S. P.}, {Garrod, R. T.}, and {Menten, K.
  M.}}}]{Belloche2016}
\bibinfo{author}{\bibnamefont{{Belloche, A.}}},
  \bibinfo{author}{\bibnamefont{{M\"uller, H. S. P.}}},
  \bibinfo{author}{\bibnamefont{{Garrod, R. T.}}}, \bibnamefont{and}
  \bibinfo{author}{\bibnamefont{{Menten, K. M.}}}, \bibinfo{journal}{Astronomy
  \& Astrophysics} \textbf{\bibinfo{volume}{587}}, \bibinfo{pages}{A91}
  (\bibinfo{year}{2016}),
  \urlprefix\url{https://doi.org/10.1051/0004-6361/201527268}.

\bibitem[{\citenamefont{{J{\o}rgensen}
  et~al.}(2016)\citenamefont{{J{\o}rgensen}, {van der Wiel}, {Coutens},
  {Lykke}, {M{\"u}ller}, {van Dishoeck}, {Calcutt}, {Bjerkeli}, {Bourke},
  {Drozdovskaya} et~al.}}]{Jorgensen2016}
\bibinfo{author}{\bibfnamefont{J.~K.} \bibnamefont{{J{\o}rgensen}}},
  \bibinfo{author}{\bibfnamefont{M.~H.~D.} \bibnamefont{{van der Wiel}}},
  \bibinfo{author}{\bibfnamefont{A.}~\bibnamefont{{Coutens}}},
  \bibinfo{author}{\bibfnamefont{J.~M.} \bibnamefont{{Lykke}}},
  \bibinfo{author}{\bibfnamefont{H.~S.~P.} \bibnamefont{{M{\"u}ller}}},
  \bibinfo{author}{\bibfnamefont{E.~F.} \bibnamefont{{van Dishoeck}}},
  \bibinfo{author}{\bibfnamefont{H.}~\bibnamefont{{Calcutt}}},
  \bibinfo{author}{\bibfnamefont{P.}~\bibnamefont{{Bjerkeli}}},
  \bibinfo{author}{\bibfnamefont{T.~L.} \bibnamefont{{Bourke}}},
  \bibinfo{author}{\bibfnamefont{M.~N.} \bibnamefont{{Drozdovskaya}}},
  \bibnamefont{et~al.}, \bibinfo{journal}{Astronomy \& Astrophysics}
  \textbf{\bibinfo{volume}{595}}, \bibinfo{eid}{A117} (\bibinfo{year}{2016}),
  \eprint{1607.08733}.

\bibitem[{\citenamefont{McGuire}(2022)}]{McGuire2022}
\bibinfo{author}{\bibfnamefont{B.~A.} \bibnamefont{McGuire}},
  \bibinfo{journal}{The Astrophysical Journal Supplement Series}
  \textbf{\bibinfo{volume}{259}}, \bibinfo{pages}{30} (\bibinfo{year}{2022}),
  \urlprefix\url{https://doi.org/10.3847/1538-4365/ac2a48}.

\bibitem[{\citenamefont{{Prasad} and {Huntress}}(1980)}]{Prasad:1980}
\bibinfo{author}{\bibfnamefont{S.~S.} \bibnamefont{{Prasad}}} \bibnamefont{and}
  \bibinfo{author}{\bibfnamefont{J.}~\bibnamefont{{Huntress}},
  \bibfnamefont{W.~T.}}, \bibinfo{journal}{Astrophysical Journal}
  \textbf{\bibinfo{volume}{43}}, \bibinfo{pages}{1} (\bibinfo{year}{1980}).

\bibitem[{\citenamefont{Shannon et~al.}(2013)\citenamefont{Shannon, Blitz,
  Goddard, and Heard}}]{Shannon2013}
\bibinfo{author}{\bibfnamefont{R.}~\bibnamefont{Shannon}},
  \bibinfo{author}{\bibfnamefont{M.}~\bibnamefont{Blitz}},
  \bibinfo{author}{\bibfnamefont{A.}~\bibnamefont{Goddard}}, \bibnamefont{and}
  \bibinfo{author}{\bibfnamefont{D.}~\bibnamefont{Heard}},
  \bibinfo{journal}{Nature Chemistry} \textbf{\bibinfo{volume}{5}},
  \bibinfo{pages}{745 } (\bibinfo{year}{2013}), \bibinfo{note}{{\copyright}
  2013 Macmillan Publishers Limited. All rights reserved. This is an author
  produced version of a paper published in Nature Chemistry. Uploaded in
  accordance with the publisher's self-archiving policy},
  \urlprefix\url{https://eprints.whiterose.ac.uk/87629/}.

\bibitem[{\citenamefont{de~la Concepci{\'{o}}n et~al.}(2021)\citenamefont{de~la
  Concepci{\'{o}}n, Puzzarini, Barone, Jim{\'{e}}nez-Serra, and
  Roncero}}]{Garcia2021}
\bibinfo{author}{\bibfnamefont{J.~G.} \bibnamefont{de~la Concepci{\'{o}}n}},
  \bibinfo{author}{\bibfnamefont{C.}~\bibnamefont{Puzzarini}},
  \bibinfo{author}{\bibfnamefont{V.}~\bibnamefont{Barone}},
  \bibinfo{author}{\bibfnamefont{I.}~\bibnamefont{Jim{\'{e}}nez-Serra}},
  \bibnamefont{and} \bibinfo{author}{\bibfnamefont{O.}~\bibnamefont{Roncero}},
  \bibinfo{journal}{The Astrophysical Journal} \textbf{\bibinfo{volume}{922}},
  \bibinfo{pages}{169} (\bibinfo{year}{2021}),
  \urlprefix\url{https://doi.org/10.3847/1538-4357/ac1e94}.

\bibitem[{\citenamefont{Farmer et~al.}(1986)\citenamefont{Farmer, Kauffman, and
  Packard}}]{farmer_autocatalytic_1986}
\bibinfo{author}{\bibfnamefont{J.}~\bibnamefont{Farmer}},
  \bibinfo{author}{\bibfnamefont{S.~A.} \bibnamefont{Kauffman}},
  \bibnamefont{and} \bibinfo{author}{\bibfnamefont{N.~H.}
  \bibnamefont{Packard}}, \bibinfo{journal}{Physica D: Nonlinear Phenomena}
  \textbf{\bibinfo{volume}{22}}, \bibinfo{pages}{50} (\bibinfo{year}{1986}),
  ISSN \bibinfo{issn}{01672789},
  \urlprefix\url{https://linkinghub.elsevier.com/retrieve/pii/0167278986902332}.

\bibitem[{\citenamefont{Banzhaf}(1993)}]{banzhaf_self-replicating_1993}
\bibinfo{author}{\bibfnamefont{W.}~\bibnamefont{Banzhaf}},
  \bibinfo{journal}{Biological Cybernetics} \textbf{\bibinfo{volume}{69}},
  \bibinfo{pages}{275} (\bibinfo{year}{1993}), ISSN \bibinfo{issn}{0340-1200,
  1432-0770}, \urlprefix\url{http://link.springer.com/10.1007/BF00203124}.

\bibitem[{\citenamefont{Dittrich and
  Banzhaf}(1998)}]{dittrich_self-evolution_1998}
\bibinfo{author}{\bibfnamefont{P.}~\bibnamefont{Dittrich}} \bibnamefont{and}
  \bibinfo{author}{\bibfnamefont{W.}~\bibnamefont{Banzhaf}},
  \bibinfo{journal}{Artificial Life} \textbf{\bibinfo{volume}{4}},
  \bibinfo{pages}{203} (\bibinfo{year}{1998}), ISSN \bibinfo{issn}{1064-5462,
  1530-9185},
  \urlprefix\url{https://direct.mit.edu/artl/article/4/2/203-220/2298}.

\bibitem[{\citenamefont{Dittrich et~al.}(2000)\citenamefont{Dittrich, Liljeros,
  Soulier, and Banzhaf}}]{dittrich_spontaneous_2000}
\bibinfo{author}{\bibfnamefont{P.}~\bibnamefont{Dittrich}},
  \bibinfo{author}{\bibfnamefont{F.}~\bibnamefont{Liljeros}},
  \bibinfo{author}{\bibfnamefont{A.}~\bibnamefont{Soulier}}, \bibnamefont{and}
  \bibinfo{author}{\bibfnamefont{W.}~\bibnamefont{Banzhaf}},
  \bibinfo{journal}{Physical Review Letters} \textbf{\bibinfo{volume}{84}},
  \bibinfo{pages}{3205} (\bibinfo{year}{2000}), ISSN \bibinfo{issn}{0031-9007,
  1079-7114},
  \urlprefix\url{https://link.aps.org/doi/10.1103/PhysRevLett.84.3205}.

\bibitem[{\citenamefont{Holdship et~al.}(2017)\citenamefont{Holdship, Viti,
  Jim{\'{e}}nez-Serra, Makrymallis, and Priestley}}]{Holdship2017}
\bibinfo{author}{\bibfnamefont{J.}~\bibnamefont{Holdship}},
  \bibinfo{author}{\bibfnamefont{S.}~\bibnamefont{Viti}},
  \bibinfo{author}{\bibfnamefont{I.}~\bibnamefont{Jim{\'{e}}nez-Serra}},
  \bibinfo{author}{\bibfnamefont{A.}~\bibnamefont{Makrymallis}},
  \bibnamefont{and}
  \bibinfo{author}{\bibfnamefont{F.}~\bibnamefont{Priestley}},
  \textbf{\bibinfo{volume}{154}}, \bibinfo{pages}{38} (\bibinfo{year}{2017}),
  \urlprefix\url{https://doi.org/10.3847/1538-3881/aa773f}.

\end{thebibliography}

 \end{document}